\documentclass[%
 aip,
 jmp,%
 amsmath,amssymb,
 reprint,%
]{revtex4}

\usepackage{graphicx}
\usepackage{dcolumn}
\usepackage{bm}

\begin{document}

\title{Tunable magnetic properties in van der Waals crystals (Fe$_{1-x}$Co$_x$)$_5$GeTe$_2$}

\author{Congkuan Tian}
\author{Feihao Pan}
\author{Sheng Xu}
\author{Kun Ai}
\author{Tianlong Xia}
\author{Peng Cheng}%
\email{pcheng@ruc.edu.cn} \affiliation{Department of Physics and
Beijing Key Laboratory of Opto-electronic Functional Materials
$\&$ Micro-nano Devices, Renmin University of China, Beijing
100872, P. R. China}

\date{\today}

\begin{abstract}
We report the doping effects of cobalt on van der Waals (vdW)
magnet Fe$_5$GeTe$_2$. A series of (Fe$_{1-x}$Co$_x$)$_5$GeTe$_2$
(0$\leq$x$\leq$0.44) single crystals have been successfully grown,
their structural, magnetic and transport properties are
investigated. For x=0.20, The Curie temperature $T_C$ increases
from 276~K to 337~K. Moreover, the magnetic easy-axis is
reoriented to the $ab$-plane from the $c$-axis in undoped
Fe$_5$GeTe$_2$ with largely enhanced magnetic anisotropy. These
magnetic properties would make (Fe$_{0.8}$Co$_{0.2}$)$_5$GeTe$_2$
more effective in stabilizing magnetic order in the
two-dimensional limit. A complex magnetic phase diagram is
identified on the higher doping side. The x=0.44 crystal first
orders ferromagnetically at $T_C$=363~K then undergoes an
antiferromagnetic transition at $T_N$=335~K. Furthermore
magnetic-field-induced spin-flop transitions are observed for the
AFM ground state. Our work reveals (Fe$_{1-x}$Co$_x$)$_5$GeTe$_2$
as promising candidates for developing new spin-related
applications and proposes a method to engineer the magnetic
properties of vdW magnet.

\end{abstract}

\maketitle

The discovery of gate-tunable room-temperature ferromagnetism in
two-dimensional (2D) van der Waals (vdW) metal Fe$_3$GeTe$_2$
(FGT) has drawn a great deal of
attention\cite{2018Nature,Fei2018}. FGT has presented plenty of
novel properties which may favor applications in spintronic and
other technologies such as current-driven magnetization
switching\cite{Switch1,Switch2}, tunneling
magnetoresistance\cite{MR}, large anomalous Hall effect\cite{AHE}
and magnetic skyrmions\cite{Sky1}. Recently an analog compound
Fe$_5$GeTe$_2$ was reported with ferromagnetic (FM) behavior in
both bulk crystals and exfoliated thin flakes\cite{Fe51,Fe52}.
Comparing with FGT, bulk Fe$_5$GeTe$_2$ crystal has higher Curie
temperature $T_C$ but very small magnetic
anisotropy\cite{Fe52,Fe53}. On the other hand, it also exhibits a
magnetoelastic coupled first-order transition below
120~K\cite{Fe52}. For potential 2D magnetic materials, strong
magnetocrystalline anisotropy and high Curie temperature are both
crucial in stabilizing the long-range FM order in
monolayer-samples and developing spintronic
devices\cite{2018Nature,holedoping}. Therefore it would be
important to check whether improved magnetic properties of
Fe$_5$GeTe$_2$ could be obtained via chemical substitution, which
has been proved to be an effective way to manipulate magnetization
in ferromagnets\cite{Fe51,2018Ni,TCK,holedoping,Yang}.

In this letter, we report the successful growth and physical
properties of (Fe$_{1-x}$Co$_x$)$_5$GeTe$_2$ (0$\leq$x$\leq$0.44)
bulk single crystals. Comparing with undoped sample, 20$\%$ of Co
doping could enhance both the Curie temperature $T_C$ and magnetic
anisotropy. On the higher doping side, new antiferromagnetic (AFM)
ground states and magnetic field induced spin-flop transitions are
observed. These findings suggest Co-doped Fe$_5$GeTe$_2$ single
crystals could have promising applications in spintronic devices.

\begin{figure}[htbp]
\centering
\includegraphics[width=0.48\textwidth]{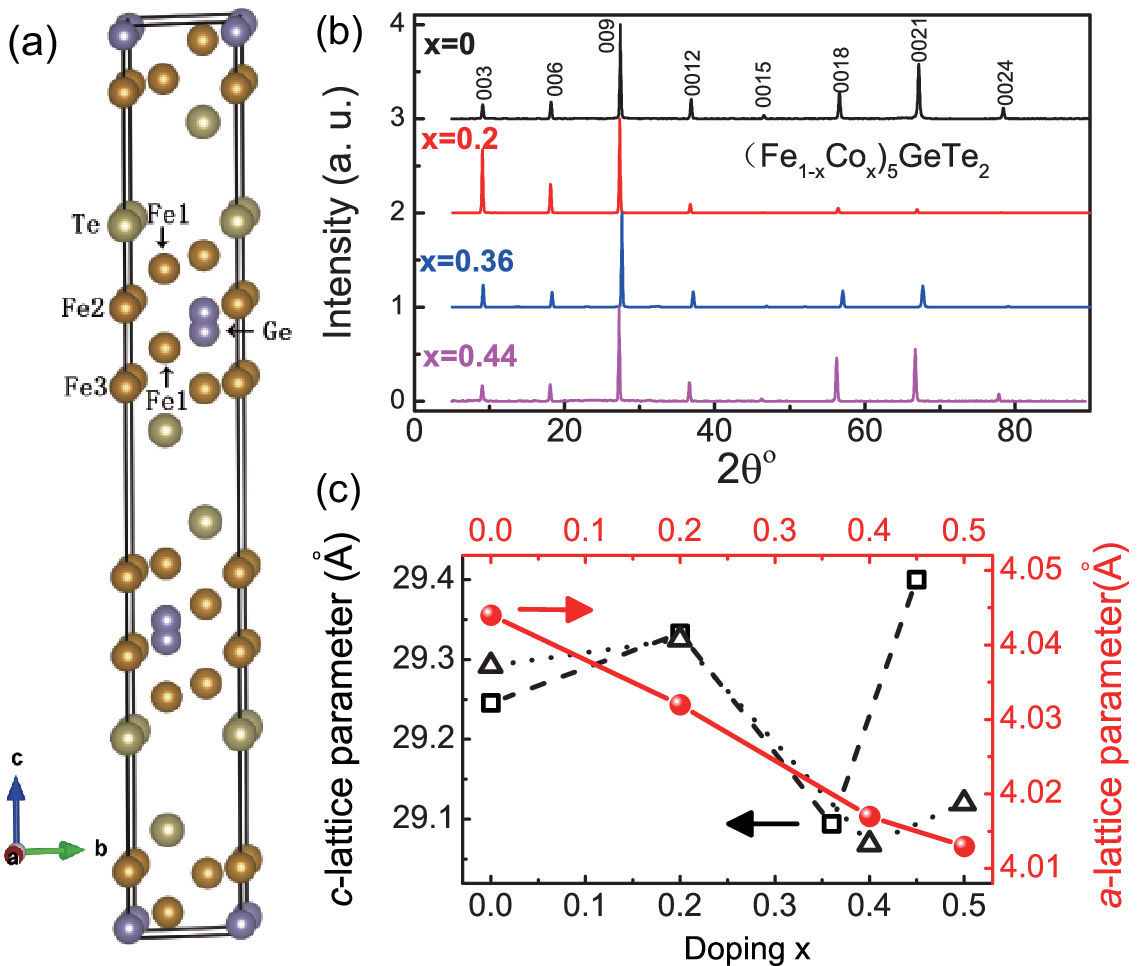}
\caption{(a) Crystal structure of Fe$_5$GeTe$_2$ with one unit
cell is shown and outlined. (b) The XRD patterns of typical
(Fe$_{1-x}$Co$_x$)$_5$GeTe$_2$ single crystals at room
temperature. (c) Doping dependent lattice parameters for
(Fe$_{1-x}$Co$_x$)$_5$GeTe$_2$ at room temperature. Solid spheres
and open triangular symbols represent the polycrystalline data.
Open squares represent the single-crystal data.}
\end{figure}

Single crystals of (Fe$_{1-x}$Co$_x$)$_5$GeTe$_2$ were grown by
the chemical vapor transport (CVT) method with iodine as the
transport agent, similiar as growing FGT in previous
work\cite{TCK}. The crystals are flat with typical dimensions of
3~mm$\times$4~mm$\times$0.1~mm and maximum doping up to x=0.44.
Attempts of crystal growth for higher Co composition were
unsuccessful. For single-crystal samples, $x$ represents actual
doping levels determined via energy dispersive x-ray spectroscopy
(EDS, Oxford X-Max 50). Some polycrystalline samples of
(Fe$_{1-x}$Co$_x$)$_5$GeTe$_2$ were synthesized by solid-state
reaction method for x-ray diffraction (XRD) studies to check the
doping evolution of $a$-lattice constants. The XRD patterns of all
samples were collected from a Bruker D8 Advance x-ray
diffractometer using Cu K$_\alpha$ radiation. The magnetization
measurements were performed using a Quantum Design MPMS3 and
resistivity measurements were carried out on a Quantum Design
physical property measurement system (QD PPMS-14T).

\begin{figure}[htbp]
\centering
\includegraphics[width=\textwidth]{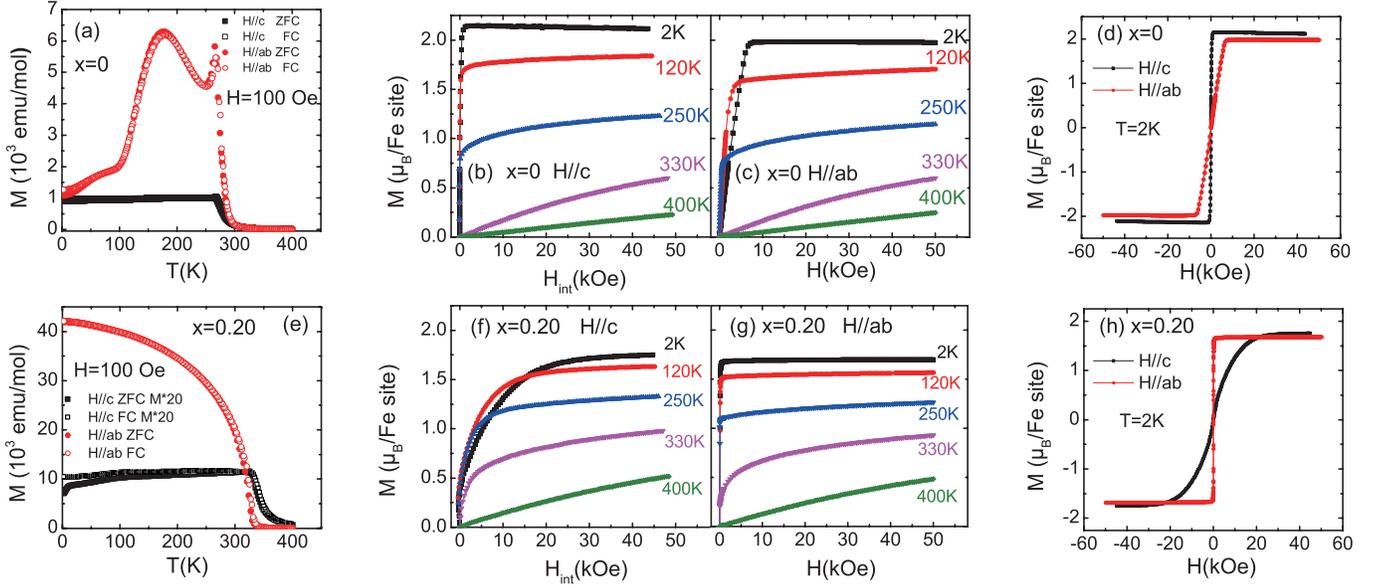}
\caption{Anisotropic magnetization data for x=0 [(a)-(d)] and
x=0.20 [(e)-(h)] single crystals respectively.
Temperature-dependent magnetization for fixed applied field are
shown in (a) and (e), isothermal magnetization curves are shown in
(b), (c), (f) and (g), magnetization hysteresis loops at 2~K are
presented in (d) and (h).}
\end{figure}

As shown in Fig. 1(a), the structure of Fe$_5$GeTe$_2$ is made up
of 2D slabs of Fe and Ge between layers of Te. The vdW gap between
Te layers makes crystals cleavable. All the XRD patterns of our
(Fe$_{1-x}$Co$_x$)$_5$GeTe$_2$ samples could be described by a
rhombohedral $R\bar{3}m$ structure as in previous
report\cite{Fe52,Fe53}. This crystal structure contains three
different Fe positions per unit cell. Based on chemical analysis
via EDS, the samples all exhibit some iron deficiency in the range
of $0.02\leq \delta \leq0.08$. Exactly speaking, the samples
should be expressed as (Fe$_{1-x-\delta}$Co$_x$)$_5$GeTe$_2$.
Nevertheless we found that the tunable magnetic properties mostly
depend on the relative cobalt doping concentration. So a
simplified expression (Fe$_{1-x}$Co$_x$)$_5$GeTe$_2$ is used to
describe our samples in this work. Fig. 1(b) presents the XRD data
of single crystals with x=0, x=0.20, x=0.36 and x=0.44
respectively. The peaks can be indexed by (0~0~3L) with
$L=1,2,3\cdots$ and no impurity peaks are found within the
instrument resolution. The doping dependent lattice parameters
derived from the x-ray data are presented in Fig. 1(c). The
$a$-lattice parameters (solid spheres) decrease monotonically with
increasing $x$ while the $c$-lattice parameters show a
nonmonotonical behavior with doping for both polycrystalline
samples (open triangular symbols) and single crystals (open
squares).

Fig. 2 shows the comparison of DC magnetization for x=0 and x=0.20
single crystals. For undoped Fe$_5$GeTe$_2$, $T_C$ is determined
to be 276~K from the temperature dependent magnetization data
(Fig. 2(a)). The soft ferromagnetic properties with low coercive
field and the magnetic remanence to saturated magnetization ratio
are similar as previous reports\cite{Fe51,Fe52,Fe53} (Fig. 2(d)).
Isothermal magnetization curves (M-H) measured under magnetic
field applied either parallel to the $c$-axis ($H$$\parallel$$c$)
or to the $ab$ plane ($H$$\parallel$$ab$) are shown in Fig. 2(b)
and (c). Demagnetization corrections have been applied on the
$H$$\parallel$$c$ data and H$_{int}$ is the internal field.
Similar as previous reports, the magnetic moments prefer to align
along the $c$-axis with an anisotropy field of 0.7~T at
2~K\cite{Fe52,Fe53}. For x=0.20, contrasting magnetic properties
are presented in Fig. 2(e)-(h). First of all, the Curie
temperature $T_C$ increases to 337~K. This value is double
confirmed from the derivatives of both $T$-dependent DC (Fig.
2(e)) and AC susceptibilities. This result is quite unusual
because typically the introduction of dopant tends to suppress
$T_C$ because of impurity-induced disorder effect as in Co-doped
Fe$_3$GeTe$_2$\cite{TCK}. Secondly, the isothermal magnetization
and hysteresis loops data show that the magnetic easy-axis of
x=0.20 crystal is reoriented to the $ab$-plane in contrast to the
$c$-axis in x=0. The easy-axis magnetism of x=0.20 has an
anisotropy field on the order of 2~T at 2~K which is also much
larger than that in x=0. Especially at high temperatures such as
120~K and 250~K, when the magnetization of Fe$_5$GeTe$_2$ becomes
nearly isotropic the x=0.20 sample still keeps a large anisotropy
field. To sum up, 20$\%$ of Co doping could effectively tune the
magnetic easy-axis of Fe$_5$GeTe$_2$, both T$_C$ and magnetic
anisotropy are significantly enhanced.

\begin{figure}[htbp]
\centering
\includegraphics[width=\textwidth]{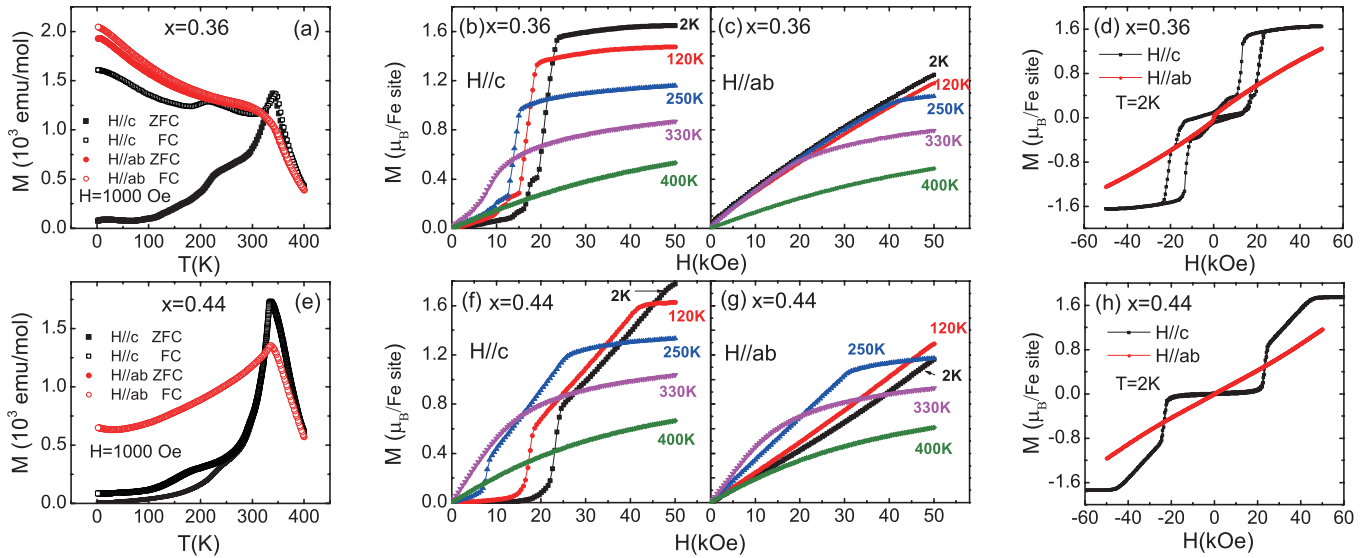}
\caption{Anisotropic magnetization data for x=0.36 [(a)-(d)] and
x=0.44 [(e)-(h)] single crystals respectively.
Temperature-dependent magnetization for fixed applied field are
shown in (a) and (e), isothermal magnetization curves are shown in
(b), (c), (f) and (g), magnetization hysteresis loops at 2~K are
presented in (d) and (h).}
\end{figure}

Another feature of x=0.20 is the absence of the phase transition
near 120~K. Previous neutron diffraction and M\"{o}ssbauer spectra
studies confirm the existence of a first-order magneto-structural
transition below 120~K for Fe$_5$GeTe$_2$ due to the FM ordering
of Fe(1) sites\cite{Fe52}. This phase transition results in
kink-features in both M-T curve (Fig. 2(a)) and temperature
dependent resistivity data (R-T curve, Fig. 4(a))\cite{Fe53}.
However for x=0.20, both the kink-features in M-T and R-T curves
disappear as shown in Fig. 2(e) and Fig. 4(a), indicating the
suppression of this phase transition by Co doping. Since the
occupancy rate of Fe(1) is just around 20$\%$\cite{Fe52}, it is
likely that the dopants mainly occupy the Fe(1) sites thus
suppressing the transition. There is also a possibility that the
enhanced $T_C$ and magnetic anisotropy are associated with the
suppression of this phase transition.

On the higher doping side, (Fe$_{1-x}$Co$_x$)$_5$GeTe$_2$ shows a
markedly different magnetic behavior. For x=0.36, the peak at
$T$=340~K in both zero-field-cooling (ZFC) and field-cooling (FC)
M-T curves with $H$$\parallel$$c$ suggest the occurrence of an
AFM-like transition (Fig. 3(a)). Fig. 3(b) shows the isothermal
magnetization data along $H$$\parallel$$c$, the virgin
magnetization curve of T=2~K begins with a gradual increase of the
magnetization then a sudden jump at around 2~T, finally
approaching the saturation magnetization. This indicates a
possible field-induced spin-flop transition from AFM to FM state.
The hysteresis loops in Fig. 3(d) suggest the FM state is much
weakened but still exists. The x=0.36 sample is more like an
intermediate phase between FM and AFM region in the phase diagram
of (Fe$_{1-x}$Co$_x$)$_5$GeTe$_2$, the magnetization data for
x=0.44 presented below provide firm evidence for the existence of
new AFM and spin-flop transitions.

\begin{figure}[htbp]
\centering
\includegraphics[width=0.48\textwidth]{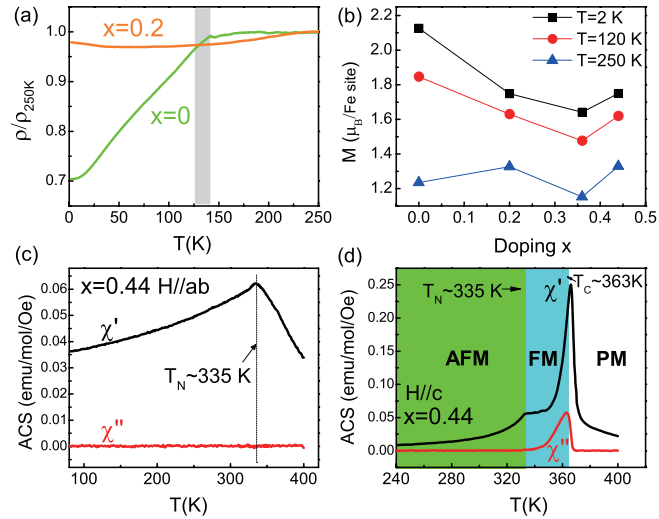}
\caption{(a) Temperature dependence of normalized electrical
resistivity measured in the $ab$ plane for x=0 and x=0.20. (b)
Doping dependence of saturated magnetic moment per Fe site at 2~K,
120~K and 250~K respectively. The temperature dependent ac
susceptibilities of x=0.44 crystal measured under oscillated AC
field of 2.0 Oe applied along $H$$\parallel$$ab$ (c) and
$H$$\parallel$$c$ directions (d). A phase diagram is superimposed
on (d).}
\end{figure}

For x=0.44, the ZFC and FC M-T curves along both directions all
exhibit sharp peak at $T_N$=335~K and drop rapidly with decreasing
temperature (Fig. 3(e)). The peak along $H$$\parallel$$c$ is much
sharper, indicative of an AFM transition with the magnetic moments
being aligned parallel to the $c$-axis. For the isothermal
magnetization and hysteresis loop along $H$$\parallel$$c$ below
250~K (Fig. 3(f) and Fig. 3(h)), the initial magnetization shows a
very weak linear increase versus field which is consistent with an
AFM state. Then at $\mu_0H\approx2.3~T$ ($T$=2~K), a steep
magnetization jump is observed, showing a typical spin-flop
transition. In general, the spin-flop transition appears at a
critical field $H_{SF}$ and the gain of magnetic energy
overcompensates the anisotropy energy required for deviation of
spin moments from the preferred orientation\cite{MA1}. Therefore
the anisotropy energy $K$ can be estimated by the equation
$K(T)=0.5(H_{SF})^2[\chi_{\perp}-\chi_{\parallel}]$\cite{MA1},
where $\chi_{\perp}$ and $\chi_{\parallel}$ are the
susceptibilities along $H$$\parallel$$ab$ and $H$$\parallel$$c$
respectively. Our calculations yield an result of
$K$=1.41$\times$10$^6$ ergs/cm$^3$ at $T$=2~K.

The saturated magnetization along $H$$\parallel$$c$ occurs at
around 5~T indicating the arrival of a complete FM state after the
spin-flop transition for x=0.44 (Fig. 3(h)). The M-H curves along
$H$$\parallel$$ab$ show a linear increase as a function of field
below 250~K indicating a paramagnetic (PM) response in PM or AFM
states (Fig. 3(g)). However the M-H curves above 330~K in both
directions exhibit FM-like non-linear behaviors suggest an
possible FM order in these temperatures. In order to further
elucidate the magnetic states for x=0.44, AC susceptibility
measurements are carried out under zero DC field and oscillated AC
field of 2.0~Oe. The data along $H$$\parallel$$ab$ is shown in
Fig. 4(c), the peak in real part $\chi'$ at 335~K and the absence
of any peaks in imaginary part $\chi''$ provide further evidences
for the AFM transition. Fig. 4(d) presents the data along
$H$$\parallel$$c$, a kink in $\chi'$ and the absence of any peaks
in $\chi''$ at the same $T_N$=335~K are consistent with the AFM
transition. Furthermore a notable peak in $\chi''$ at higher
temperature $T_C$=363~K is observed. Combining with the rapid
increase of magnetization at around 363~K in M-T curves (Fig.
3(e)) and the FM-like non-linear behavior in the M-H curves above
330~K (Fig. 3(f) and (g)), all the experimental observations
strongly suggest that there is an FM transition at $T_C$=363~K. So
the phase diagram of x=0.44 when the magnetic field is not strong
enough to cause spin-flop could be drawn in Fig.4 (d). Both DC and
AC magnetization data fully support this phase diagram.

The AFM ground state for x=0.44 could possibly be associated with
the reduction of $a$-lattice parameter. For one thing, from the FM
state in x=0 and x=0.20 to the AFM state in x=0.36 and x=0.44, the
$c$-lattice parameters change nonmonotonically. So the change of
interlayer distance along $c$-axis does not seem to be the reason
for the occurrence of AFM interactions. For the other, a previous
work shows that Ni-doped Fe$_5$GeTe$_2$ samples are all
ferromagnets with simialr $c$ but much larger $a$ parameters
comparing with our AFM ordered samples\cite{Fe51}. So the reduced
intralayer Fe-Fe distance is likely to make AFM interactions more
favorable. As in previous theoretical calculations on CrS$_2$ and
CrI$_3$\cite{JiWei1,JiWei2}, charge carrier doping or
doping-induced new stacking orders could also be possible
explanations for this tunable magnetism. So our results provide
clues and possible methods to manipulate
(Fe$_{1-x}$Co$_x$)$_5$GeTe$_2$ few layers transiting between AFM
and FM states, which could be employed in future magnetic data
recording and information processing. As to the magnetic structure
of x=0.44, our magnetization data reveal that the magnetic moments
tend to align along the $c$-axis, but could not give whether
intralayer or interlayer AFM (as CrI$_3$\cite{CrI3}) order is
preferred. Further neutron scattering investigations are needed
for the final solution. In the development of new spin-related
applications, either intrinsic 2D FM or AFM ground state is
useful\cite{RMP,Review}. So x=0.20 and x=0.44 single crystals
could all serve as new vdW candidate magnets for designing
nanoscale spintronic devices. Especially the saturated magnetic
moments of (Fe$_{1-x}$Co$_x$)$_5$GeTe$_2$ do not have much
reduction with increasing $x$ and the values at $T$=250~K are even
doping independent as shown in Fig. 4(b). This is also good news
for applications.

To summarize our results, we have discovered that the doping of
cobalt could significantly tune the bulk magnetic properties of 2D
vdW magnets (Fe$_{1-x}$Co$_x$)$_5$GeTe$_2$. For x=0.20, both the
Curie temperature and the magnetic anisotropy have increased
greatly which makes (Fe$_{0.8}$Co$_{0.2}$)$_5$GeTe$_2$ a better
potential room-temperature intrinsic 2D magnet comparing with the
undoped parent compound. For x=0.44, a new AFM ground state with
$T_N$=335~K is identified below the FM transition at $T_C$=363~K.
Magnetic field induced spin-flop transitions are also observed in
this doping range. Our results provide new vdW candidate materials
which could potentially be exfoliated for functional vdW
heterostructures and devices.

\section{Acknowledgments}
The authors thank the helpful discussion with Prof. Wei Ji. This
work is supported by the National Natural Science Foundation of
China (No. 11227906 and No. 11204373).

\bibliography{Bibtex}

\begin{thebibliography}{20}
\expandafter\ifx\csname natexlab\endcsname\relax\def\natexlab#1{#1}\fi
\expandafter\ifx\csname bibnamefont\endcsname\relax
  \def\bibnamefont#1{#1}\fi
\expandafter\ifx\csname bibfnamefont\endcsname\relax
  \def\bibfnamefont#1{#1}\fi
\expandafter\ifx\csname citenamefont\endcsname\relax
  \def\citenamefont#1{#1}\fi
\expandafter\ifx\csname url\endcsname\relax
  \def\url#1{\texttt{#1}}\fi
\expandafter\ifx\csname urlprefix\endcsname\relax\def\urlprefix{URL }\fi
\providecommand{\bibinfo}[2]{#2}
\providecommand{\eprint}[2][]{\url{#2}}

\bibitem[{\citenamefont{Deng et~al.}(2018)\citenamefont{Deng, Yu, Song, Zhang,
  Wang, Sun, Yi, Wu, Wu, Zhu et~al.}}]{2018Nature}
\bibinfo{author}{\bibfnamefont{Y.}~\bibnamefont{Deng}},
  \bibinfo{author}{\bibfnamefont{Y.}~\bibnamefont{Yu}},
  \bibinfo{author}{\bibfnamefont{Y.}~\bibnamefont{Song}},
  \bibinfo{author}{\bibfnamefont{J.}~\bibnamefont{Zhang}},
  \bibinfo{author}{\bibfnamefont{N.~Z.} \bibnamefont{Wang}},
  \bibinfo{author}{\bibfnamefont{Z.}~\bibnamefont{Sun}},
  \bibinfo{author}{\bibfnamefont{Y.}~\bibnamefont{Yi}},
  \bibinfo{author}{\bibfnamefont{Y.~Z.} \bibnamefont{Wu}},
  \bibinfo{author}{\bibfnamefont{S.}~\bibnamefont{Wu}},
  \bibinfo{author}{\bibfnamefont{J.}~\bibnamefont{Zhu}}, \bibnamefont{et~al.},
  \bibinfo{journal}{Nature} \textbf{\bibinfo{volume}{563}}, \bibinfo{pages}{94}
  (\bibinfo{year}{2018}), ISSN \bibinfo{issn}{1476-4687},
  \urlprefix\url{https://doi.org/10.1038/s41586-018-0626-9}.

\bibitem[{\citenamefont{Fei et~al.}(2018)\citenamefont{Fei, Huang, Malinowski,
  Wang, Song, Sanchez, Yao, Xiao, Zhu, May et~al.}}]{Fei2018}
\bibinfo{author}{\bibfnamefont{Z.}~\bibnamefont{Fei}},
  \bibinfo{author}{\bibfnamefont{B.}~\bibnamefont{Huang}},
  \bibinfo{author}{\bibfnamefont{P.}~\bibnamefont{Malinowski}},
  \bibinfo{author}{\bibfnamefont{W.}~\bibnamefont{Wang}},
  \bibinfo{author}{\bibfnamefont{T.}~\bibnamefont{Song}},
  \bibinfo{author}{\bibfnamefont{J.}~\bibnamefont{Sanchez}},
  \bibinfo{author}{\bibfnamefont{W.}~\bibnamefont{Yao}},
  \bibinfo{author}{\bibfnamefont{D.}~\bibnamefont{Xiao}},
  \bibinfo{author}{\bibfnamefont{X.}~\bibnamefont{Zhu}},
  \bibinfo{author}{\bibfnamefont{A.~F.} \bibnamefont{May}},
  \bibnamefont{et~al.}, \bibinfo{journal}{Nature Materials}
  \textbf{\bibinfo{volume}{17}}, \bibinfo{pages}{778} (\bibinfo{year}{2018}),
  ISSN \bibinfo{issn}{1476-4660},
  \urlprefix\url{https://doi.org/10.1038/s41563-018-0149-7}.

\bibitem[{\citenamefont{Wang et~al.}(2019)\citenamefont{Wang, Tang, Xia, He,
  Zhang, Liu, Wan, Fang, Guo, Yang et~al.}}]{Switch1}
\bibinfo{author}{\bibfnamefont{X.}~\bibnamefont{Wang}},
  \bibinfo{author}{\bibfnamefont{J.}~\bibnamefont{Tang}},
  \bibinfo{author}{\bibfnamefont{X.}~\bibnamefont{Xia}},
  \bibinfo{author}{\bibfnamefont{C.}~\bibnamefont{He}},
  \bibinfo{author}{\bibfnamefont{J.}~\bibnamefont{Zhang}},
  \bibinfo{author}{\bibfnamefont{Y.}~\bibnamefont{Liu}},
  \bibinfo{author}{\bibfnamefont{C.}~\bibnamefont{Wan}},
  \bibinfo{author}{\bibfnamefont{C.}~\bibnamefont{Fang}},
  \bibinfo{author}{\bibfnamefont{C.}~\bibnamefont{Guo}},
  \bibinfo{author}{\bibfnamefont{W.}~\bibnamefont{Yang}}, \bibnamefont{et~al.},
  \bibinfo{journal}{Science Advances} \textbf{\bibinfo{volume}{5}}
  (\bibinfo{year}{2019}),
  \eprint{https://advances.sciencemag.org/content/5/8/eaaw8904.full.pdf},
  \urlprefix\url{https://advances.sciencemag.org/content/5/8/eaaw8904}.

\bibitem[{\citenamefont{Alghamdi et~al.}(2019)\citenamefont{Alghamdi, Lohmann,
  Li, Jothi, Shao, Aldosary, Su, Fokwa, and Shi}}]{Switch2}
\bibinfo{author}{\bibfnamefont{M.}~\bibnamefont{Alghamdi}},
  \bibinfo{author}{\bibfnamefont{M.}~\bibnamefont{Lohmann}},
  \bibinfo{author}{\bibfnamefont{J.}~\bibnamefont{Li}},
  \bibinfo{author}{\bibfnamefont{P.~R.} \bibnamefont{Jothi}},
  \bibinfo{author}{\bibfnamefont{Q.}~\bibnamefont{Shao}},
  \bibinfo{author}{\bibfnamefont{M.}~\bibnamefont{Aldosary}},
  \bibinfo{author}{\bibfnamefont{T.}~\bibnamefont{Su}},
  \bibinfo{author}{\bibfnamefont{B.~P.~T.} \bibnamefont{Fokwa}},
  \bibnamefont{and} \bibinfo{author}{\bibfnamefont{J.}~\bibnamefont{Shi}},
  \bibinfo{journal}{Nano Letters} \textbf{\bibinfo{volume}{19}},
  \bibinfo{pages}{4400} (\bibinfo{year}{2019}), \bibinfo{note}{pMID: 31177784},
  \eprint{https://doi.org/10.1021/acs.nanolett.9b01043},
  \urlprefix\url{https://doi.org/10.1021/acs.nanolett.9b01043}.

\bibitem[{\citenamefont{Wang et~al.}(2018{\natexlab{a}})\citenamefont{Wang,
  Sapkota, Taniguchi, Watanabe, Mandrus, and Morpurgo}}]{MR}
\bibinfo{author}{\bibfnamefont{Z.}~\bibnamefont{Wang}},
  \bibinfo{author}{\bibfnamefont{D.}~\bibnamefont{Sapkota}},
  \bibinfo{author}{\bibfnamefont{T.}~\bibnamefont{Taniguchi}},
  \bibinfo{author}{\bibfnamefont{K.}~\bibnamefont{Watanabe}},
  \bibinfo{author}{\bibfnamefont{D.}~\bibnamefont{Mandrus}}, \bibnamefont{and}
  \bibinfo{author}{\bibfnamefont{A.~F.} \bibnamefont{Morpurgo}},
  \bibinfo{journal}{Nano Letters} \textbf{\bibinfo{volume}{18}},
  \bibinfo{pages}{4303} (\bibinfo{year}{2018}{\natexlab{a}}),
  \bibinfo{note}{pMID: 29870263},
  \eprint{https://doi.org/10.1021/acs.nanolett.8b01278},
  \urlprefix\url{https://doi.org/10.1021/acs.nanolett.8b01278}.

\bibitem[{\citenamefont{Kim et~al.}(2018)\citenamefont{Kim, Seo, Lee, Ko, Kim,
  Jang, Ok, Lee, Jo, Kang et~al.}}]{AHE}
\bibinfo{author}{\bibfnamefont{K.}~\bibnamefont{Kim}},
  \bibinfo{author}{\bibfnamefont{J.}~\bibnamefont{Seo}},
  \bibinfo{author}{\bibfnamefont{E.}~\bibnamefont{Lee}},
  \bibinfo{author}{\bibfnamefont{K.-T.} \bibnamefont{Ko}},
  \bibinfo{author}{\bibfnamefont{B.~S.} \bibnamefont{Kim}},
  \bibinfo{author}{\bibfnamefont{B.~G.} \bibnamefont{Jang}},
  \bibinfo{author}{\bibfnamefont{J.~M.} \bibnamefont{Ok}},
  \bibinfo{author}{\bibfnamefont{J.}~\bibnamefont{Lee}},
  \bibinfo{author}{\bibfnamefont{Y.~J.} \bibnamefont{Jo}},
  \bibinfo{author}{\bibfnamefont{W.}~\bibnamefont{Kang}}, \bibnamefont{et~al.},
  \bibinfo{journal}{Nature Materials} \textbf{\bibinfo{volume}{17}},
  \bibinfo{pages}{794} (\bibinfo{year}{2018}), ISSN \bibinfo{issn}{1476-4660},
  \urlprefix\url{https://doi.org/10.1038/s41563-018-0132-3}.

\bibitem[{\citenamefont{Ding et~al.}(2020)\citenamefont{Ding, Li, Xu, Li, Hou,
  Liu, Xi, Xu, Yao, and Wang}}]{Sky1}
\bibinfo{author}{\bibfnamefont{B.}~\bibnamefont{Ding}},
  \bibinfo{author}{\bibfnamefont{Z.}~\bibnamefont{Li}},
  \bibinfo{author}{\bibfnamefont{G.}~\bibnamefont{Xu}},
  \bibinfo{author}{\bibfnamefont{H.}~\bibnamefont{Li}},
  \bibinfo{author}{\bibfnamefont{Z.}~\bibnamefont{Hou}},
  \bibinfo{author}{\bibfnamefont{E.}~\bibnamefont{Liu}},
  \bibinfo{author}{\bibfnamefont{X.}~\bibnamefont{Xi}},
  \bibinfo{author}{\bibfnamefont{F.}~\bibnamefont{Xu}},
  \bibinfo{author}{\bibfnamefont{Y.}~\bibnamefont{Yao}}, \bibnamefont{and}
  \bibinfo{author}{\bibfnamefont{W.}~\bibnamefont{Wang}},
  \bibinfo{journal}{Nano Letters} \textbf{\bibinfo{volume}{20}},
  \bibinfo{pages}{868} (\bibinfo{year}{2020}), \bibinfo{note}{pMID: 31869236},
  \eprint{https://doi.org/10.1021/acs.nanolett.9b03453},
  \urlprefix\url{https://doi.org/10.1021/acs.nanolett.9b03453}.

\bibitem[{\citenamefont{Stahl et~al.}(2018)\citenamefont{Stahl, Shlaen, and
  Johrendt}}]{Fe51}
\bibinfo{author}{\bibfnamefont{J.}~\bibnamefont{Stahl}},
  \bibinfo{author}{\bibfnamefont{E.}~\bibnamefont{Shlaen}}, \bibnamefont{and}
  \bibinfo{author}{\bibfnamefont{D.}~\bibnamefont{Johrendt}},
  \bibinfo{journal}{Z. Anorg. Allg. Chem.} \textbf{\bibinfo{volume}{644}},
  \bibinfo{pages}{1923} (\bibinfo{year}{2018}),
  \eprint{https://onlinelibrary.wiley.com/doi/pdf/10.1002/zaac.201800456},
  \urlprefix\url{https://onlinelibrary.wiley.com/doi/abs/10.1002/zaac.20180045%
6}.

\bibitem[{\citenamefont{May et~al.}(2019{\natexlab{a}})\citenamefont{May,
  Ovchinnikov, Zheng, Hermann, Calder, Huang, Fei, Liu, Xu, and
  McGuire}}]{Fe52}
\bibinfo{author}{\bibfnamefont{A.~F.} \bibnamefont{May}},
  \bibinfo{author}{\bibfnamefont{D.}~\bibnamefont{Ovchinnikov}},
  \bibinfo{author}{\bibfnamefont{Q.}~\bibnamefont{Zheng}},
  \bibinfo{author}{\bibfnamefont{R.}~\bibnamefont{Hermann}},
  \bibinfo{author}{\bibfnamefont{S.}~\bibnamefont{Calder}},
  \bibinfo{author}{\bibfnamefont{B.}~\bibnamefont{Huang}},
  \bibinfo{author}{\bibfnamefont{Z.}~\bibnamefont{Fei}},
  \bibinfo{author}{\bibfnamefont{Y.}~\bibnamefont{Liu}},
  \bibinfo{author}{\bibfnamefont{X.}~\bibnamefont{Xu}}, \bibnamefont{and}
  \bibinfo{author}{\bibfnamefont{M.~A.} \bibnamefont{McGuire}},
  \bibinfo{journal}{ACS Nano} \textbf{\bibinfo{volume}{13}},
  \bibinfo{pages}{4436} (\bibinfo{year}{2019}{\natexlab{a}}),
  \bibinfo{note}{pMID: 30865426},
  \eprint{https://doi.org/10.1021/acsnano.8b09660},
  \urlprefix\url{https://doi.org/10.1021/acsnano.8b09660}.

\bibitem[{\citenamefont{May et~al.}(2019{\natexlab{b}})\citenamefont{May,
  Bridges, and McGuire}}]{Fe53}
\bibinfo{author}{\bibfnamefont{A.~F.} \bibnamefont{May}},
  \bibinfo{author}{\bibfnamefont{C.~A.} \bibnamefont{Bridges}},
  \bibnamefont{and} \bibinfo{author}{\bibfnamefont{M.~A.}
  \bibnamefont{McGuire}}, \bibinfo{journal}{Phys. Rev. Materials}
  \textbf{\bibinfo{volume}{3}}, \bibinfo{pages}{104401}
  (\bibinfo{year}{2019}{\natexlab{b}}),
  \urlprefix\url{https://link.aps.org/doi/10.1103/PhysRevMaterials.3.104401}.

\bibitem[{\citenamefont{Park et~al.}(2020)\citenamefont{Park, Kim, Liu, Hwang,
  Kim, Kim, Kim, Petrovic, Hwang, Mo et~al.}}]{holedoping}
\bibinfo{author}{\bibfnamefont{S.~Y.} \bibnamefont{Park}},
  \bibinfo{author}{\bibfnamefont{D.~S.} \bibnamefont{Kim}},
  \bibinfo{author}{\bibfnamefont{Y.}~\bibnamefont{Liu}},
  \bibinfo{author}{\bibfnamefont{J.}~\bibnamefont{Hwang}},
  \bibinfo{author}{\bibfnamefont{Y.}~\bibnamefont{Kim}},
  \bibinfo{author}{\bibfnamefont{W.}~\bibnamefont{Kim}},
  \bibinfo{author}{\bibfnamefont{J.-Y.} \bibnamefont{Kim}},
  \bibinfo{author}{\bibfnamefont{C.}~\bibnamefont{Petrovic}},
  \bibinfo{author}{\bibfnamefont{C.}~\bibnamefont{Hwang}},
  \bibinfo{author}{\bibfnamefont{S.-K.} \bibnamefont{Mo}},
  \bibnamefont{et~al.}, \bibinfo{journal}{Nano Letters}
  \textbf{\bibinfo{volume}{20}}, \bibinfo{pages}{95} (\bibinfo{year}{2020}),
  \bibinfo{note}{pMID: 31752490},
  \eprint{https://doi.org/10.1021/acs.nanolett.9b03316},
  \urlprefix\url{https://doi.org/10.1021/acs.nanolett.9b03316}.

\bibitem[{\citenamefont{Drachuck et~al.}(2018)\citenamefont{Drachuck, Salman,
  Masters, Taufour, Lamichhane, Lin, Straszheim, Bud'ko, and
  Canfield}}]{2018Ni}
\bibinfo{author}{\bibfnamefont{G.}~\bibnamefont{Drachuck}},
  \bibinfo{author}{\bibfnamefont{Z.}~\bibnamefont{Salman}},
  \bibinfo{author}{\bibfnamefont{M.~W.} \bibnamefont{Masters}},
  \bibinfo{author}{\bibfnamefont{V.}~\bibnamefont{Taufour}},
  \bibinfo{author}{\bibfnamefont{T.~N.} \bibnamefont{Lamichhane}},
  \bibinfo{author}{\bibfnamefont{Q.}~\bibnamefont{Lin}},
  \bibinfo{author}{\bibfnamefont{W.~E.} \bibnamefont{Straszheim}},
  \bibinfo{author}{\bibfnamefont{S.~L.} \bibnamefont{Bud'ko}},
  \bibnamefont{and} \bibinfo{author}{\bibfnamefont{P.~C.}
  \bibnamefont{Canfield}}, \bibinfo{journal}{Phys. Rev. B}
  \textbf{\bibinfo{volume}{98}}, \bibinfo{pages}{144434}
  (\bibinfo{year}{2018}),
  \urlprefix\url{https://link.aps.org/doi/10.1103/PhysRevB.98.144434}.

\bibitem[{\citenamefont{Tian et~al.}(2019)\citenamefont{Tian, Wang, Ji, Wang,
  Xia, Wang, Liu, Zhang, and Cheng}}]{TCK}
\bibinfo{author}{\bibfnamefont{C.-K.} \bibnamefont{Tian}},
  \bibinfo{author}{\bibfnamefont{C.}~\bibnamefont{Wang}},
  \bibinfo{author}{\bibfnamefont{W.}~\bibnamefont{Ji}},
  \bibinfo{author}{\bibfnamefont{J.-C.} \bibnamefont{Wang}},
  \bibinfo{author}{\bibfnamefont{T.-L.} \bibnamefont{Xia}},
  \bibinfo{author}{\bibfnamefont{L.}~\bibnamefont{Wang}},
  \bibinfo{author}{\bibfnamefont{J.-J.} \bibnamefont{Liu}},
  \bibinfo{author}{\bibfnamefont{H.-X.} \bibnamefont{Zhang}}, \bibnamefont{and}
  \bibinfo{author}{\bibfnamefont{P.}~\bibnamefont{Cheng}},
  \bibinfo{journal}{Phys. Rev. B} \textbf{\bibinfo{volume}{99}},
  \bibinfo{pages}{184428} (\bibinfo{year}{2019}),
  \urlprefix\url{https://link.aps.org/doi/10.1103/PhysRevB.99.184428}.

\bibitem[{\citenamefont{Liu et~al.}(2019)\citenamefont{Liu, Wang, Pu, Zhang,
  Yang, Musho, and Chen}}]{Yang}
\bibinfo{author}{\bibfnamefont{J.}~\bibnamefont{Liu}},
  \bibinfo{author}{\bibfnamefont{A.}~\bibnamefont{Wang}},
  \bibinfo{author}{\bibfnamefont{K.}~\bibnamefont{Pu}},
  \bibinfo{author}{\bibfnamefont{S.}~\bibnamefont{Zhang}},
  \bibinfo{author}{\bibfnamefont{J.}~\bibnamefont{Yang}},
  \bibinfo{author}{\bibfnamefont{T.}~\bibnamefont{Musho}}, \bibnamefont{and}
  \bibinfo{author}{\bibfnamefont{L.}~\bibnamefont{Chen}},
  \bibinfo{journal}{Phys. Chem. Chem. Phys.} \textbf{\bibinfo{volume}{21}},
  \bibinfo{pages}{7588} (\bibinfo{year}{2019}),
  \urlprefix\url{http://dx.doi.org/10.1039/C9CP00151D}.

\bibitem[{\citenamefont{He and Ueda}(2008)}]{MA1}
\bibinfo{author}{\bibfnamefont{Z.}~\bibnamefont{He}} \bibnamefont{and}
  \bibinfo{author}{\bibfnamefont{Y.}~\bibnamefont{Ueda}},
  \bibinfo{journal}{Phys. Rev. B} \textbf{\bibinfo{volume}{77}},
  \bibinfo{pages}{052402} (\bibinfo{year}{2008}),
  \urlprefix\url{https://link.aps.org/doi/10.1103/PhysRevB.77.052402}.

\bibitem[{\citenamefont{Wang et~al.}(2018{\natexlab{b}})\citenamefont{Wang,
  Zhou, Pan, Qiao, Kong, Kaun, and Ji}}]{JiWei1}
\bibinfo{author}{\bibfnamefont{C.}~\bibnamefont{Wang}},
  \bibinfo{author}{\bibfnamefont{X.}~\bibnamefont{Zhou}},
  \bibinfo{author}{\bibfnamefont{Y.}~\bibnamefont{Pan}},
  \bibinfo{author}{\bibfnamefont{J.}~\bibnamefont{Qiao}},
  \bibinfo{author}{\bibfnamefont{X.}~\bibnamefont{Kong}},
  \bibinfo{author}{\bibfnamefont{C.-C.} \bibnamefont{Kaun}}, \bibnamefont{and}
  \bibinfo{author}{\bibfnamefont{W.}~\bibnamefont{Ji}}, \bibinfo{journal}{Phys.
  Rev. B} \textbf{\bibinfo{volume}{97}}, \bibinfo{pages}{245409}
  (\bibinfo{year}{2018}{\natexlab{b}}),
  \urlprefix\url{https://link.aps.org/doi/10.1103/PhysRevB.97.245409}.

\bibitem[{\citenamefont{Jiang et~al.}(2019)\citenamefont{Jiang, Wang, Chen,
  Zhong, Yuan, Lu, and Ji}}]{JiWei2}
\bibinfo{author}{\bibfnamefont{P.}~\bibnamefont{Jiang}},
  \bibinfo{author}{\bibfnamefont{C.}~\bibnamefont{Wang}},
  \bibinfo{author}{\bibfnamefont{D.}~\bibnamefont{Chen}},
  \bibinfo{author}{\bibfnamefont{Z.}~\bibnamefont{Zhong}},
  \bibinfo{author}{\bibfnamefont{Z.}~\bibnamefont{Yuan}},
  \bibinfo{author}{\bibfnamefont{Z.-Y.} \bibnamefont{Lu}}, \bibnamefont{and}
  \bibinfo{author}{\bibfnamefont{W.}~\bibnamefont{Ji}}, \bibinfo{journal}{Phys.
  Rev. B} \textbf{\bibinfo{volume}{99}}, \bibinfo{pages}{144401}
  (\bibinfo{year}{2019}),
  \urlprefix\url{https://link.aps.org/doi/10.1103/PhysRevB.99.144401}.

\bibitem[{\citenamefont{Huang et~al.}(2017)\citenamefont{Huang, Clark,
  Navarromoratalla, Klein, Cheng, Seyler, Zhong, Schmidgall, Mcguire, and
  Cobden}}]{CrI3}
\bibinfo{author}{\bibfnamefont{B.}~\bibnamefont{Huang}},
  \bibinfo{author}{\bibfnamefont{G.}~\bibnamefont{Clark}},
  \bibinfo{author}{\bibfnamefont{E.}~\bibnamefont{Navarromoratalla}},
  \bibinfo{author}{\bibfnamefont{D.~R.} \bibnamefont{Klein}},
  \bibinfo{author}{\bibfnamefont{R.}~\bibnamefont{Cheng}},
  \bibinfo{author}{\bibfnamefont{K.~L.} \bibnamefont{Seyler}},
  \bibinfo{author}{\bibfnamefont{D.}~\bibnamefont{Zhong}},
  \bibinfo{author}{\bibfnamefont{E.}~\bibnamefont{Schmidgall}},
  \bibinfo{author}{\bibfnamefont{M.~A.} \bibnamefont{Mcguire}},
  \bibnamefont{and} \bibinfo{author}{\bibfnamefont{D.~H.}
  \bibnamefont{Cobden}}, \bibinfo{journal}{Nature}
  \textbf{\bibinfo{volume}{546}}, \bibinfo{pages}{270} (\bibinfo{year}{2017}).

\bibitem[{\citenamefont{Baltz et~al.}(2018)\citenamefont{Baltz, Manchon, Tsoi,
  Moriyama, Ono, and Tserkovnyak}}]{RMP}
\bibinfo{author}{\bibfnamefont{V.}~\bibnamefont{Baltz}},
  \bibinfo{author}{\bibfnamefont{A.}~\bibnamefont{Manchon}},
  \bibinfo{author}{\bibfnamefont{M.}~\bibnamefont{Tsoi}},
  \bibinfo{author}{\bibfnamefont{T.}~\bibnamefont{Moriyama}},
  \bibinfo{author}{\bibfnamefont{T.}~\bibnamefont{Ono}}, \bibnamefont{and}
  \bibinfo{author}{\bibfnamefont{Y.}~\bibnamefont{Tserkovnyak}},
  \bibinfo{journal}{Rev. Mod. Phys.} \textbf{\bibinfo{volume}{90}},
  \bibinfo{pages}{015005} (\bibinfo{year}{2018}),
  \urlprefix\url{https://link.aps.org/doi/10.1103/RevModPhys.90.015005}.

\bibitem[{\citenamefont{Li et~al.}(2019)\citenamefont{Li, Ruan, and
  Zeng}}]{Review}
\bibinfo{author}{\bibfnamefont{H.}~\bibnamefont{Li}},
  \bibinfo{author}{\bibfnamefont{S.}~\bibnamefont{Ruan}}, \bibnamefont{and}
  \bibinfo{author}{\bibfnamefont{Y.-J.} \bibnamefont{Zeng}},
  \bibinfo{journal}{Advanced Materials} \textbf{\bibinfo{volume}{31}},
  \bibinfo{pages}{1900065} (\bibinfo{year}{2019}),
  \eprint{https://onlinelibrary.wiley.com/doi/pdf/10.1002/adma.201900065},
  \urlprefix\url{https://onlinelibrary.wiley.com/doi/abs/10.1002/adma.20190006%
5}.

\end{thebibliography}

\end{document}